\documentclass[superscriptaddress,showpacs,aps,twocolumn,prb]{revtex4}

\usepackage{graphicx}
\usepackage{amssymb}

\def\fref#1{Fig.~\ref{#1}}
\renewcommand{\vec}{\mathbf}

\begin{document}

\title{Suppression of magnetic relaxation processes in melt-textured
YBa$_2$Cu$_3$O$_x$ superconductors by a transverse ac magnetic field}

\author{L.~M. Fisher}
\affiliation{All-Russian Electrical Engineering Institute, 12 Krasnokazarmennaya Street, 111250
Moscow, Russia}
\author{A.~V. Kalinov}
\affiliation{All-Russian Electrical Engineering Institute, 12 Krasnokazarmennaya Street, 111250
Moscow, Russia}
\author{I.~F. Voloshin}
\affiliation{All-Russian Electrical Engineering Institute, 12 Krasnokazarmennaya Street, 111250
Moscow, Russia}
\author{V.~A. Yampol'skii}
\affiliation{Institute for Radiophysics and Electronics NASU, 12 Proskura Street, 61085 Kharkov,
Ukraine}

\begin{abstract}
The effect of the transverse AC magnetic field on relaxation process in YBa$_2$Cu$_3$O$_x$
melt-textured superconductor was studied. A factor of 50 suppression of the relaxation rate could
be achieved at the expense of some reduction in the maximum trapped field, with the
magnetic-induction gradient being unchanged. This phenomenon is interpreted as a result of an
increase of the pinning force after the action of the transverse AC magnetic field that is
confirmed by the measurement of the trapped-induction distribution.
\end{abstract}

\date{\today}


\maketitle

The electrodynamics of hard superconductors was under study for a long time. A wide variety of
interesting phenomena were discovered and interpreted on the basis of the developed concepts. One
of them is the suppression of static magnetization under the action of a transverse AC magnetic
field. First, this phenomenon was observed and interpreted by Yamafuji and
coauthors.\cite{yama1,yama2}. If one places a superconducting strip into a perpendicular static
(DC) magnetic field ${\vec H}$ and then apply an AC magnetic field ${\vec h}(t)$ along the strip
surface, the critical profile of the static magnetic flux changes noticeably. The short vortices
oriented in perpendicular to the strip are shown to bend and move in such a way that the static
magnetic flux tends to become homogeneous. This phenomenon and its consequences were studied
theoretically in detail in recent papers.\cite{brand1, brand2}

The other origin of the suppression of the static magnetization by the transverse AC magnetic field
(the collapse phenomenon) was considered in Refs. \onlinecite{col-new, cut, kin, cross, felpap}. If
the vortex length $L$ exceeds significantly the penetration depth of the AC field, the vortex bend
does not play an essential role and the flux-line cutting mechanism (the phenomenon predicted by
Clem \cite{Clem}) is put in the forefront. As was shown in Refs.~\onlinecite{cut, kin, felpap}, it
is the flux-line cutting that provides the homogenization of the static magnetic flux in the wide
areas of the sample bulk where the AC field has penetrated (the creation of collapse zone). If the
AC amplitude $h$ exceeds the penetration field $H_p$ the distribution of the static magnetic
induction becomes homogeneous in the whole sample bulk and the magnetic moment
disappears.\cite{cross,Willemin}

It is well-known that the inhomogeneous magnetic flux distribution is metastable. According to the
classical paper by Anderson,\cite{Ander} such a state of a superconductor relaxes to the
homogeneous one following the logarithmic law. Within the existing concept on the collapse, one
could expect a noticeable decrease of the relaxation rate of the magnetic moment after the action
of the AC field. Indeed, the surface homogeneous region of the sample, the collapse zone, should be
filled by the vortices and the critical gradient should be established before the vortices start to
leave the sample and the magnetic moment begins to decrease. In the present paper we have checked
this assumption and made sure that the relaxation rate of the DC magnetic moment is actually
decreased significantly by the action of the transverse AC magnetic field. Surprisingly, we have
observed a striking concomitant phenomenon which appears to be of general interest in vortex
matter. Not only the magnetic moment does not changes, but the spatial distribution of the static
magnetic flux holds the shape without significant relaxation for a long time after the action of AC
field. Moreover, this distribution conserves after the moderate temperature increase. This strongly
suggests that the critical current density grows after the action.

Melt-textured YBCO plate-like samples of $9.3\times7.4\times1.5$~mm$^3$ (sample A) and
$8\times6.3\times1.5$~mm$^3$ (sample B) in sizes were cut from a homogeneous part of bulk textured
cylinders grown by the seeding technique. The homogeneity was checked by a scanning Hall-probe. The
\textbf{c}-axis was perpendicular to the largest face of the sample. At $T = 77$~K, the
characteristic value of the critical current density $J_c$ in the \textbf{ab} plane is of the order
of 13 and 22~kA/cm$^2$ for samples A and B (in self-field), respectively. The `static'
magnetization $M$ was measured by a vibrating sample magnetometer in the external magnetic field
$\mathbf{H}\|\mathbf{c}$ created by an electromagnet. The zero-field cooled sample was exposed to
the magnetic field of 12~kOe which was further reduced to 5~kOe; from this point, the relaxation
measurement was started. The above-mentioned fields are essentially higher than the penetration
field of the sample (1.6~kOe). A commercial Hall-probe was fixed on the sample to measure the
evolution of the magnetic induction locally. The other Hall-probe with the sensitive zone of
$0.3\times0.3$~mm$^2$ was used to scan the magnetic induction distribution on the sample surface.
The distance from the sample to the sensitive zone of the probe was about 0.2~mm. The Hall
measurements were performed in the zero field after exposition of the sample to $H=12$~kOe. The AC
magnetic field $\mathbf{h}\|\mathbf{ab}$ was a computer-generated triangle-wave with the frequency
$F=140$~Hz. The AC field could be applied for a definite number of periods and stopped exactly at
the end of the period (at $h=0$). In our experiments we used 999 full cycles of the AC field. Most
of the measurements were performed at the liquid nitrogen temperature $T=77$~K.

\begin{figure}[tb]
\includegraphics{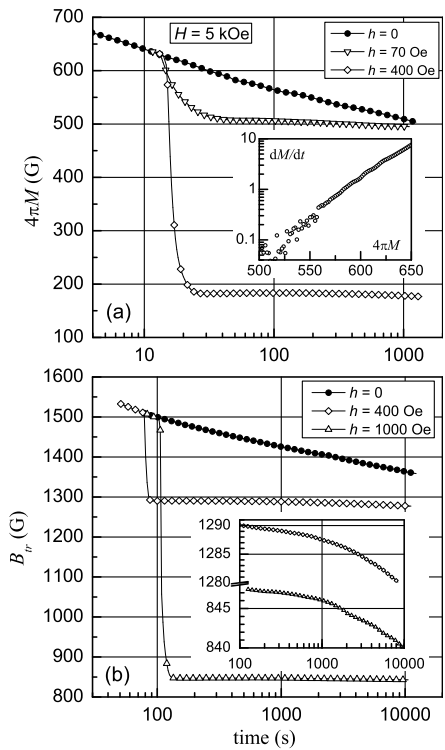}
\caption{\label{relax} Relaxation of the magnetization $M$ (a) and trapped magnetic induction
$B_{tr}$ (b) without (closed symbols) and with the action of the AC field (open symbols) for sample
A. The upper inset shows the dependence of $\mathrm d M/\mathrm d t \propto E$ on $M\propto J$; the
lower shows the tails of suppressed relaxation in enlarged scale.}
% ($\square$) ({\Large $\circ$}) ($\vartriangle$) ($\triangledown$) ($\blacksquare$)
\end{figure}

Figure \ref{relax}(a) demonstrates the influence of the orthogonal AC magnetic field on the
relaxation of $M$. The conventional (without the AC field) relaxation is shown to follow the
logarithmic law, that implies an exponential current-voltage characteristic (CVC) $E \propto
\exp[(-U/kT)(1-J/J_{c0})]$ (see inset), where $J_{c0}$ is a depinning current at the zero
temperature and $U$ is a pinning-well depth.\cite{Ander} For the other run, at $t\approx 20$~s the
AC field of the amplitude $h$ was applied, followed by the sharp drop in the magnetization due to
the collapse-effect.\cite{cross} When the AC field action has finished, we can see almost no
relaxation for the first 100 seconds and essentially reduced logarithmic relaxation rate $S =
\mathrm d M/ \mathrm d \ln t$ for the rest of the observation period.

% FIG.1

To reduce the minor effect of the magnetic-field relaxation in the electromagnet on the
magnetization we studied the temporal dependence of the trapped magnetic induction $B_{tr}$ in the
central part of the sample without any external magnetic field by the Hall-probe. Three curves in
\fref{relax}(b) correspond to the same magnetic prehistory of the sample, but differ from each
other because of the action of the AC field with different amplitudes. The upper curve is obtained
for the case when the AC field was not switched on during the measurement at all, $h=0$. Two lower
curves demonstrate the influence of the AC field on the relaxation of $B_{tr}$. All curves in
\fref{relax}(b) follow the same logarithmic-like law before switching on the AC field. Switching on
the AC field results in a giant suppression of the relaxation rate (5 -- 60 times). The inset to
\fref{relax}(b) shows that the decrease of the relaxation rate depends essentially on the amplitude
of the AC field and $B_{tr}$ does not follow the logarithmic law for the whole time-window.

The abrupt decrease of the static magnetization under the action of the AC field (the jumps shown
in \fref{relax}) is connected to the suppression of the DC shielding currents in all sample regions
where the AC field penetrates. The suppression of the relaxation of $M$ and $B_{tr}$ could be
easily interpreted as a result of the collapse: vortices can not leave the sample before the
sufficient gradient of the magnetic induction would be restored in the surface regions of the
sample. From the macroscopic point of view, the other explanation looks plausible. As we reduce the
magnetization (and current density) significantly, we should get a huge decrease in $\mathrm d
M/\mathrm d t\propto E$ in accordance with the exponential CVC (see inset to \fref{relax}(a)).
However, this explanation will work only in the case of the uniform distribution of the shielding
currents. As we will see below this is not the case.

\begin{figure}[tb]
\includegraphics{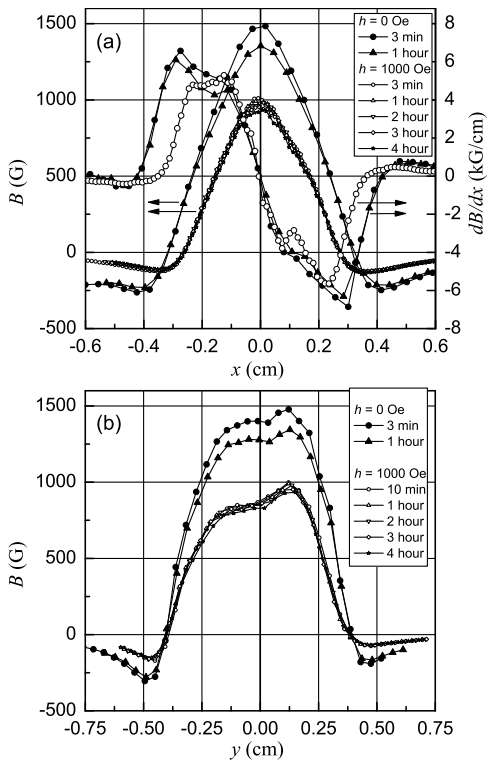}
\caption{\label{B-x} Relaxation of the magnetic-induction distribution along the $x$ (a) and $y$
(b) directions on the surface of sample A without (closed symbols) and after the action of the AC
field (open symbols). Panel (a) (the right scale) also shows derivatives $\mathrm d B/\mathrm d
x$.}
\end{figure}

\begin{figure}[tb]
\includegraphics{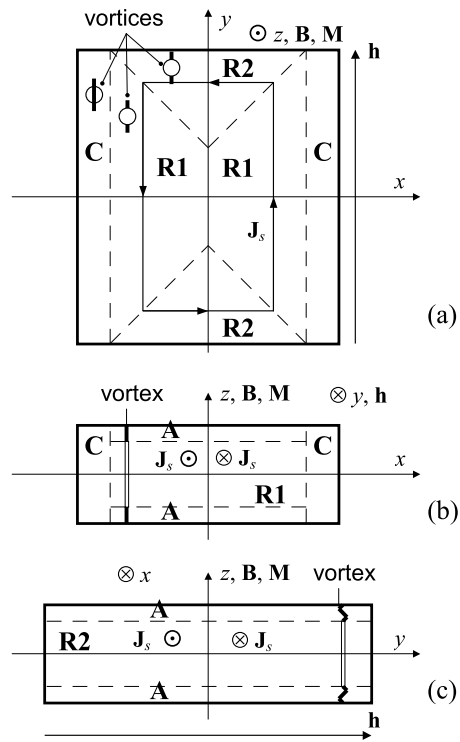}
\caption{\label{sketch} Schematic view of the cross-sections of the sample. Regions marked with `C'
are fully penetrated by the AC field, there is no shielding current ($J_s$) in the $xy$-plane here.
The vortices in regions R1 and R2 are partially disturbed (bent or entangled in the $zy$-plane).
This disturbance occurs in the A-regions corresponding to the AC field penetration depth along the
$z$-axis.}
\end{figure}

To clarify the real influence of the AC field on the relaxation process we have investigated the
temporal evolution of the DC magnetic field distribution by the scanning Hall-probe. The spatial
distribution $B(x)$ (across the AC field direction) of the trapped DC magnetic field is shown in
\fref{B-x}(a). The unperturbed curves (closed symbols) demonstrate Bean-like profiles of the
magnetic flux distribution which relaxes obviously in one hour. After the action of the AC field
(open symbols) the relaxation is suppressed significantly in accordance with the data of
\fref{relax}(b). The change in the profile width along $x$-axis is due to the collapse of the DC
shielding currents in the regions of the sample where the AC field has penetrated (the `collapsed'
regions marked with `C' in \fref{sketch}). Even at a first glance at the curves in \fref{B-x}(a)
one can reveal a very important consequence of the action of the AC field. The magnetic field
distribution in the central part of the sample before and after this action differs by the vertical
displacement, with the gradient being the same. Numerically calculated $\mathrm{d}B/\mathrm{d}x$
curves shown in \fref{B-x} (a), confirm this observation to be true everywhere except most outer
regions. This means that the current densities in R1-regions remain their initial values in all
sample regions where the AC field has not penetrated along the $x$-direction. However the regions
occupied by the currents flowing along the $x$-direction (R2) are not changed. This, together with
the decrease of the maximum trapped field results in the decrease of the current density in R2
regions (see \fref{B-x}(b)) that makes them subcritical and suppresses the relaxation in the
$y$-direction. This decrease of the local induction itself could affect the relaxation rate $S$
directly, due to an increase of the $J_c$ or a change in the CVC exponent. However, the change of
the CVC exponent with the magnetic field is too low to explain the huge change in $S$.\cite{Kupfer}
The influence of local $B$ on $S$ through the $J_c$ change could be estimated comparing the data of
\fref{relax} (a) and (b). The external magnetic field in \fref{relax} (a) is well higher than the
penetration field and the ac field. In this case, the change of the local $B$ is essentially lower
than for zero-field measurement \fref{relax} (b), but the effect of the relaxation suppression does
exist and is qualitatively the same.

% FIG.2

The most interesting feature of the relaxation after the action of the AC field is the following.
Not only the magnetization and the magnetic induction but the distribution of the DC-field
gradient, i.\ e.\ distribution of the DC shielding currents, stops the relaxation after the action
of the AC field. The vortices in the central part of the sample (R1 and R2-regions in
\fref{sketch}) could not move into the collapsed regions (`C'). This is probably due to
impossibility to overcome the cutting barrier,\cite{park,Indenbom} because the C-regions are (at
least partially) filled with the vortices aligned with the $y$-axis. However, the other possible
origin of the suppression of relaxation has to be considered.

% FIG.3

We suppose the penetration of the AC field from the largest ($xy$) faces of the sample to be of
great importance. The first possible reason is an increase in the vortex length accompanied by the
pinning energy gain, tending to the increase of the $J_c$, because the vortices turn out to be
`anchored' by their tails (in the A-regions). The elongation of the vortices occurs in such a way
(along $y$-axis) that there is no additional Lorentz force directed outside the sample (along $x$-
and $y$-directions). Thus, the shielding currents become subcritical resulting in the exponential
decrease of the relaxation rate. However, to explain the significant decrease of $S$ by the
vortex-length increase along, the relative vortex elongation must be of the order of the
logarithmic relaxation-rate decrease, which is hardly possible. Moreover, in the collective pinning
approach \cite{Blatter} the vortex elongation gives only sublinear term in the free energy. The
other explanation could be based on the anisotropy of the pinning force in the YBCO
superconductors. After the action of the AC field, parts (segments) of the vortices in the
A-regions can be aligned and locked-in in the \textbf{ab}-plane leading to an increase of the
pinning force. The action of the AC field could also result in the bend or entangling of the
vortices in the A-regions at the scale of 100 -- 1000 vortex-lattice periods. As a result, a
significant increase of the vortex-bundle size can take place. This, in turn, can lead to the
increase of the collective pinning potential \cite{Blatter} and to the observed suppression of the
relaxation rate.

% FIG.4
\begin{figure}[tb]
\includegraphics{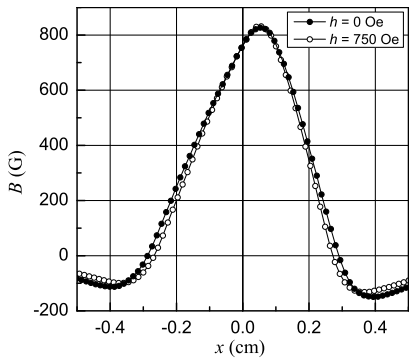}
\caption{\label{enhance} Distribution of the trapped induction for sample B obtained without
(closed symbols) and after the action of the AC field (open symbols). The magneto-thermal history
is described in the text.}
\end{figure}

To check if the action of the AC field  does increase the critical current density, the following
experiment was done. Suggesting the shielding currents becomes subcritical after the action we
could increase the temperature without decrease of the current density. Our scanning-Hall-probe
setup permits only 77~K measurements, so we have to decrease the temperature back to 77~K to
investigate the changes. This temperature decrease should not change the magnetic flux profile in
any way in the absence of the external magnetic field, thus we can study a distribution frozen from
the higher temperature. The magneto-thermal history was as follows. Sample B was cooled to 83~K in
the external magnetic field of 10~kOe then the field was switched off. At this point the AC field
was applied (or not applied in the other run). Further, the sample was heated by 3~K to `develop' a
difference between perturbed and non-perturbed flux profiles. Before this heating, they should look
like the profiles on \fref{B-x}(a) (the gradients are the same). Upon heating the non-perturbed
profile should follow the $J_c$ changes and perturbed one will go from subcritical to a critical
state (possibly with a higher $J_c$). Finally, the sample was cooled down to 77~K and profiles were
measured. (Note, that these profiles correspond to the critical current at $T=86$~K.) These
experiments were repeated several times, with the temperatures reproducibility being better than
0.05~K. The result is shown in \fref{enhance}. The perturbed profile is seen to keep narrower form
but the maximum field is the same. This means that the magnetic-induction distribution is steeper
and the critical current density in the `perturbed' sample is higher. The crude evaluation gives
the $J_c$ increase of at least 8\%.

Summarizing, we have observed the giant decrease of the relaxation rate after the action of the
transverse AC magnetic field. This effect exists in the relaxation of the `static' magnetization
and the local value of the trapped magnetic induction. After this action, the distribution of the
magnetic induction across the AC field direction turns out to be frozen with the same current
density. This distribution conserves even if the temperature of the sample increased by several
Kelvin giving a new critical state with a higher $J_c$. We suppose the increase of the vortex
length accompanied by partial locking of the vortex segments in the \textbf{ab}-planes could be a
reason of this phenomenon. The observed effect can also be a result of the increase of the
collective pinning potential due to the increase of the vortex-bundles dimension in the
\textbf{ab}-plane.

We thank A. L. Rakhmanov for helpful discussions. This work is supported by INTAS (grant 01-2282),
RFBR (grant 03-02-17169), and Russian National Program on Superconductivity
(contract~40.012.1.1.11.46).

\end{document}